\documentclass[12pt]{iopart}

\usepackage[english]{babel}
\usepackage[utf8x]{inputenc}
\usepackage[T1]{fontenc}

\usepackage{iopams}
\usepackage{epigraph}
\usepackage{float}
\usepackage{subcaption}
\usepackage{graphicx}
\usepackage{float}

\usepackage{bm}

\usepackage{xcolor}

\usepackage{tikz}
\usetikzlibrary{calc}

\usepackage[a4paper,top=3cm,bottom=2cm,left=3cm,right=3cm,marginparwidth=1.75cm]{geometry}

\usepackage{graphicx}
\usepackage[colorlinks=true, allcolors=blue]{hyperref}
\usepackage{cancel}

\usepackage{cite}

\newcommand{\bbm}{\begin{multline}}
\newcommand{\eem}{\end{multline}}
\newcommand{\be}{\begin{equation}}
\newcommand{\ee}{\end{equation}}
\newcommand{\bea}{\begin{eqnarray}}
\newcommand{\eea}{\end{eqnarray}}

\begin{document}


\title[The density profile of a Coulomb plasma on a cylinder]{The density profile of a Coulomb plasma on a cylinder: boundary oscillations}

 \author{Gabriel Cardoso}
 \address{Tsung-Dao Lee Institute, Shanghai Jiao Tong University, Shanghai, 201210, China}

 \author{Jean-Marie St\'ephan}
 \address{CNRS, ENS de Lyon, LPENSL, UMR5672, 69342, Lyon cedex 07, France}

 \author{Alexander G. Abanov}
 \address{Department of Physics and Astronomy, Stony Brook University, Stony Brook, NY 11794, USA}

\date{}                     


\begin{abstract}
We present Monte Carlo simulations of the two-dimensional one-component plasma (2D OCP) confined to a cylindrical geometry, focusing on density profiles, fluctuations, and their connection to bulk correlation functions. The cylindrical geometry eliminates geometric frustration, allowing for a precise study of boundary density oscillations, the dependence on boundary conditions, and their relationship to the melting transition and triangular lattice structure. By triangulating particle configurations, we quantify the exponential suppression of topological defects in the crystalline phase. Furthermore, we propose an oriented correlation function that better links boundary density profiles with bulk correlation functions, motivating anisotropic generalizations of the phase-field crystal (PFC) model. These results provide new insights into the interplay between boundary effects, bulk correlations, and phase transitions in the 2D OCP.

\end{abstract}

\section{Introduction}
 \label{sec:intro}

The two-dimensional one-component plasma (2D OCP) is a classical statistical mechanics model, consisting of charged particles that interact via a logarithmic Coulomb potential in the presence of a neutralizing background. This system is relevant to a wide range of fields, including condensed matter physics, astrophysics, and plasma physics \cite{1980-BausHansen-PhysRep, caillol1982monte, deLeeuw}. In particular, it plays an important role in the study of the fractional quantum Hall effect (FQHE) \cite{laughlin1983anomalous,moosavi2024quantum, cappelli2021w, wiegmann2013hydrodynamics}. The partition function of the 2D OCP coincides with the eigenvalue distributions of certain random matrix ensembles \cite{ginibre1965statistical, forrester2010log,forrester2024dualities, akemann2024complex}, further contributing to its enduring mathematical interest \cite{forrester2016analogies, serfaty2024lectures, ameur2023planar}.

At low temperatures, the 2D OCP undergoes Wigner crystallization, where particles arrange into a triangular lattice to minimize their potential energy \cite{alastuey1981classical, choquard1983cooperative, Tkachenko:1966ux}. While this ordered phase has been extensively studied, the nature of the melting transition at $\Gamma_m \approx 140$ remains an open question. Proposed scenarios range from a weakly first-order transition to the Berezinskii-Kosterlitz-Thouless-Halperin-Nelson-Young (BKTHNY) mechanism, which involves an intermediate hexatic phase \cite{berezinsky1970destruction, kosterlitz1973ordering, halperin1978theory, kleinert1989gauge, khrapak2018note}. The unique role of logarithmic interactions, as compared to shorter-ranged potentials, adds an additional layer of complexity to these questions \cite{caillol1982monte, kapfer2015two, radloff1984freezing}.

Of particular interest is the boundary density profile of the 2D OCP, which undergoes a transition at $\Gamma=2$ from monotonic decay to oscillatory behavior \cite{jancovici1981exact, jancovici1982classical1, badiali1983surface, can2014edgelaughlin, can2015exact, datta1996edge}. This phenomenon has been studied using various analytical and numerical methods, including large-$N$ expansions, Monte Carlo simulations, and exact solutions \cite{morf1986monte, jancovici1982classical2, levesque2000charge, zabrodin2006large,mcclarty2007}. Recent studies have further advanced this understanding by investigating boundary fluctuations, overcrowding probabilities, and density oscillations in related systems \cite{thoma2024overcrowding, ameur2022two, ji2024universal, estienne2022cornering}. 

In \cite{cardoso2020boundary}, we argued that the damped oscillations in the edge density profile can be understood as the formation of a crystalline layer, or freezing, at the edge. In this paper, we build on this quantitative picture through Monte Carlo simulations of the OCP on the cylinder. On the cylinder, the boundary of the plasma-occupied region has vanishing geodesic curvature, which prevents the formation of extrinsic topological defects. In other words, the cylindrical geometry is fully compatible with a perfect triangular lattice without disclinations. Topological defects appear only when they are favored by free energy, a phenomenon we study in this work. We also take advantage of this feature to explore different boundary conditions and to separate the averaging over angular orientations of the density profile. Furthermore, we connect bulk density correlations to boundary density profiles using the phase-field crystal method. These analyses provide a more precise correspondence between density profiles, correlations, and the thermodynamics of the 2D OCP.

The paper is organized as follows. In Section \ref{sec:cyl}, we review the properties of the Coulomb plasma and the triangular lattice on the cylinder. In Section \ref{sec:MC}, we describe our Monte Carlo simulation method. Section \ref{sec:profiles} presents our results for the density profiles at different temperatures obtained from the simulations. We discuss the independence of these results on boundary conditions, including a comparison with our previous simulations \cite{cardoso2020boundary} on the disk geometry. We also investigate the density of topological defects at very low temperatures and the stripe regime observed in the two-dimensional densities. In Section \ref{sec:correlations}, we explore the connection between density profiles and correlation functions using the phase-field crystal method. Additionally, we propose a calculation based on orientational correlations computed from the bulk. Finally, in Section \ref{sec:discussion}, we summarize our main conclusions and discuss potential directions for future research.

\section{Coulomb plasma on the cylinder}
 \label{sec:cyl}

\begin{figure}
\centering
\includegraphics[width=0.8\linewidth]{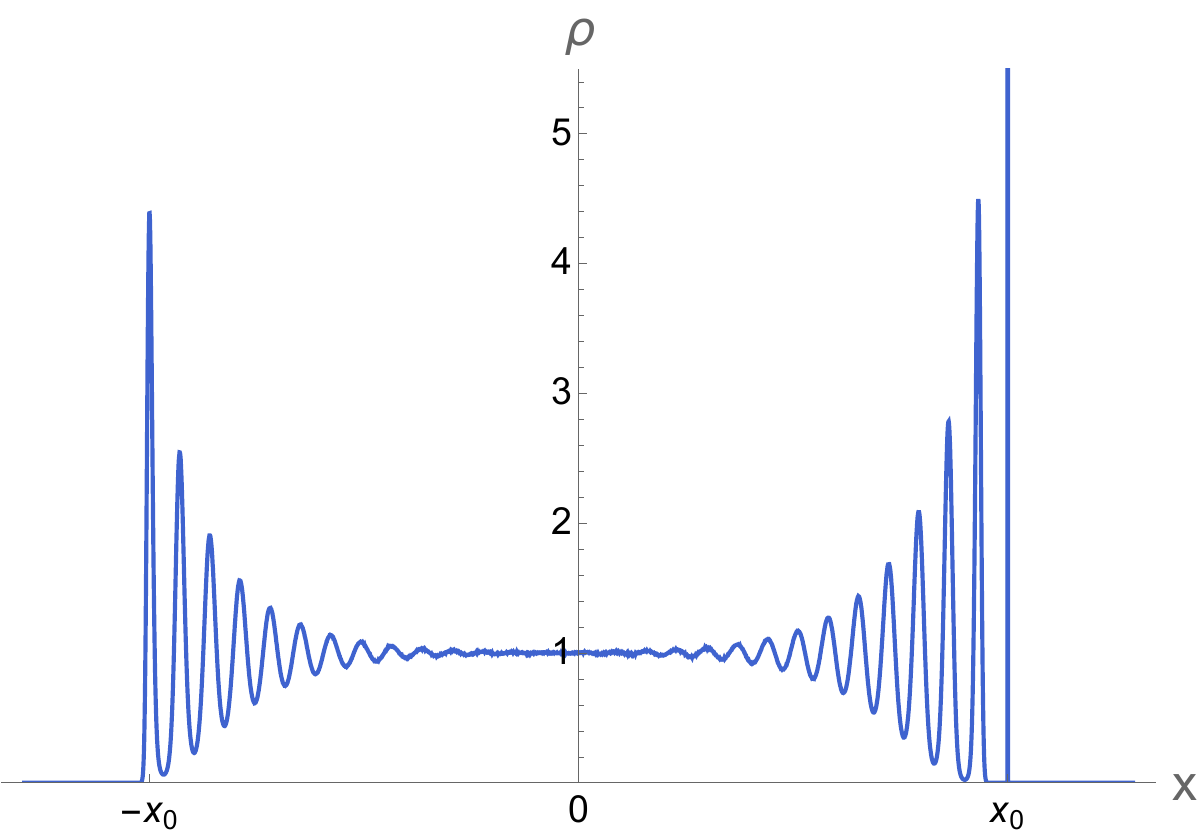}
\caption{Density profile of the 2D OCP on the cylinder at inverse temperature $\Gamma=100$ with $N=900$ particles. In the bulk, the density reaches a constant value of $\rho_0=1$, while strong oscillations appear near the boundaries. These oscillations are associated with the structure of the triangular lattice on the cylinder, from which the size of the plasma droplet, $2x_0$, can be computed (\ref{eq:x0lattice}). This plot shows mixed boundary conditions: particle positions are fixed on the right boundary at $x=x_0$, while they freely fluctuate on the left boundary at $x=-x_0$.}
\label{fig:density130}
\end{figure}

The one-component plasma (OCP) consists of $N$ particles, each with the same charge $q$, in a homogeneous neutralizing background. The potential energy of a given configuration is
\begin{equation}\label{eq:energy}
    E = q^2\sum_{1\leq j<k\leq N}v(\mathbf{x}_j,\mathbf{x}_k) + q^2\sum_{j=1}^N W(\mathbf{x}_j)\,,
\end{equation}
where the Coulomb interaction is defined as the Green's function of the Laplace-Beltrami operator:
\begin{equation}
    \Delta v(\mathbf{x},\mathbf{x}') = -2\pi\delta(\mathbf{x}-\mathbf{x}'),
\end{equation}
and the background potential $q W(\mathbf{x})$ satisfies
\begin{equation}
    \Delta W(\mathbf{x}) = 2\pi \rho_0,
\end{equation}
corresponding to the background charge density $-q\rho_0$. In the following, we fix the length units by setting $\rho_0=1$.

We consider the two-dimensional OCP on a cylinder of radius $R$, where the Coulomb potential is given by
\begin{equation}
    v(\mathbf{x},\mathbf{x'}) = -\ln\left[2\cosh\left(\frac{x-x'}{R}\right) - 2\cos\left(\frac{y-y'}{R}\right)\right]^{1/2},
 \label{intcyl}
\end{equation}
where $\mathbf{x}=(x,y)$, $y$ parametrizes the circumference, with $y \sim y + 2\pi R$, and the background potential is
\begin{equation}
    W(\mathbf{x}) = \pi x^2\,.
 \label{Wcyl}
\end{equation}
We study this system in the canonical ensemble,
\begin{equation}
    Z = \int \prod_{j=1}^N d^2 \mathbf{x}_j \; \exp\left\{-\Gamma \left[\sum_{1\leq j<k\leq N}v(\mathbf{x}_j,\mathbf{x}_k) +  \sum_{j=1}^N  W(\mathbf{x}_j)\right]\right\}\,,
 \label{Zcyl}
\end{equation}
where $\Gamma = \frac{q^2}{k_B T}$ is the dimensionless inverse temperature. The system is in the liquid phase for $\Gamma < \Gamma_m \approx 140$ and forms a triangular lattice for $\Gamma > \Gamma_m$. The particle density is given by the expectation value
\begin{equation}
    \rho(\mathbf{x}) = \left\langle\sum_{i=1}^N \delta(\mathbf{x}_i - \mathbf{x})\right\rangle_\Gamma.
\end{equation}
In the liquid phase ($\Gamma < \Gamma_m$), and under boundary conditions that preserve the rotational symmetry of the cylinder (i.e., translations in the $y$ direction), $\rho(\mathbf{x}) = \rho(x)$ represents a one-dimensional density profile. We also consider boundary conditions that preserve only discrete translations in the $y$ direction. In this case, the continuous symmetry persists in the bulk, and the density profile $\rho(x)$ can be obtained by averaging over the $y$-direction.

For $N \to \infty$, the plasma extends over the entire cylinder, with a constant density fixed by the charge screening condition,
\begin{equation}
    \rho(x) = \rho_0.
\end{equation}
At finite $N$, a boundary forms between the bulk region, where $\rho(x) \to \rho_0$, and the external region, where $\rho(x) \to 0$. In the weak-coupling regime ($\Gamma \ll 1$), the density profile near the boundary is well-described by Debye-H\"uckel mean-field theory. However, at strong coupling, its shape becomes highly nontrivial, as shown in Figure \ref{fig:density130}. The most striking feature is the appearance of strong oscillations near the boundary. As previously reported \cite{zabrodin2006large, can2014singular}, the boundary features are highly singular in the large-$N$ limit, making them difficult to capture with approximate mean-field theory methods. In our previous work \cite{cardoso2020boundary}, we demonstrated that these features are more clearly understood in terms of the neighboring crystalline phase. On the cylinder, this picture can be made more precise due to the simpler structure of the triangular lattice.

\subsection{Triangular lattice on the cylinder}
 \label{sec:triangle}

\begin{figure}[h]
\begin{subfigure}{.48\linewidth}
\centering
\includegraphics[width=\linewidth]{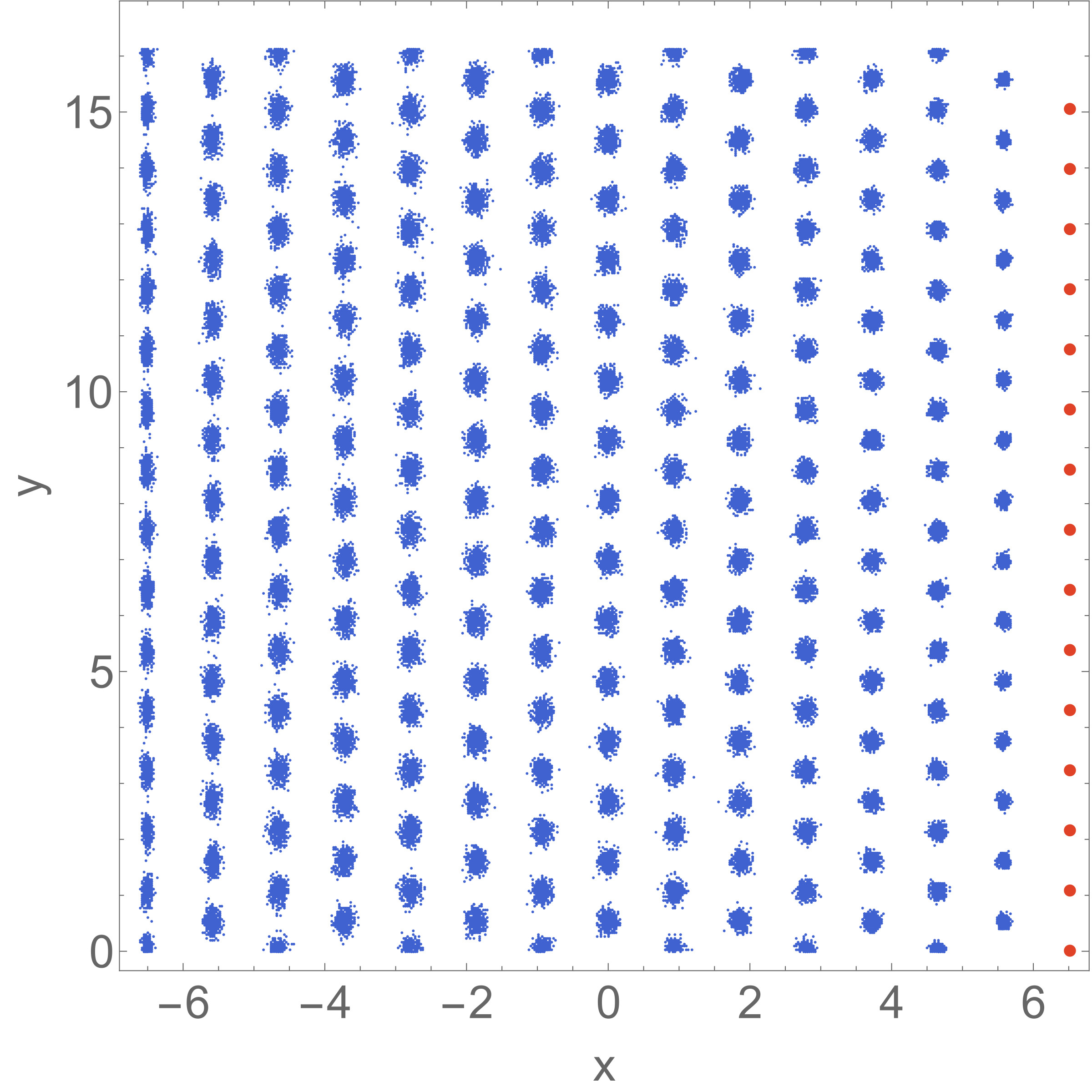}
\end{subfigure}
\hfill
\begin{subfigure}{.48\linewidth}
\centering
\includegraphics[width=\linewidth]{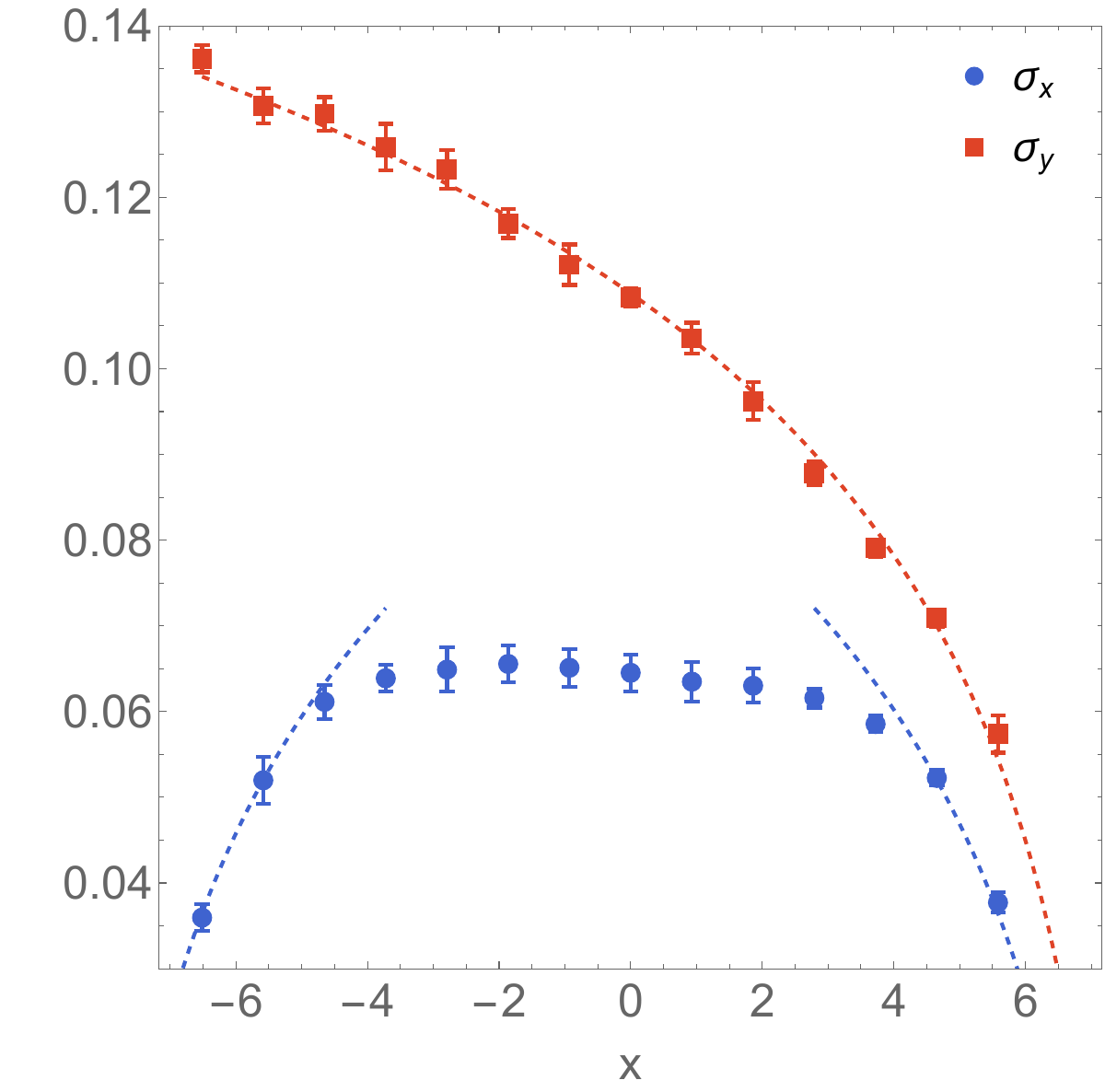}
\end{subfigure}
\caption{Triangular lattice and particle position fluctuations. Left: A total of $400$ superimposed Monte Carlo samples for the plasma at $\Gamma=400 > \Gamma_m$ with $N=225$ particles. At these low temperatures, the particles form a perfect triangular lattice. The mixed boundary conditions are also shown: the rightmost lattice plane is fixed (red dots), while at the left boundary, the particles can freely fluctuate. Right: The variance of particle position fluctuations along the $x$- and $y$-directions as a function of $x$. These fluctuations vanish at the right boundary and increase logarithmically with distance (dashed lines). At the left boundary, only fluctuations in the $x$-direction vanish, as the boundary conditions there are  set by the external potential.}
\label{fig:lattice}
\end{figure}

In the planar case, with a rotationally symmetric background potential, a finite number of particles will occupy a disk of radius $R = \sqrt{N}/\pi$. However, it is generally not possible to form a perfect triangular lattice on the disk. This geometric frustration leads to the presence of disclination defects even at zero temperature, where they are not thermodynamically favorable. In the cylindrical geometry, this obstruction can be overcome. 

Consider a configuration of $N = m^2$ charges. The triangular lattice spacing is given by
\begin{equation}
    a_0 = \left(\frac{4}{3}\right)^{1/4},
\end{equation}
and we fix the radius of the cylinder to
\begin{equation}
    R = \frac{m a_0}{2\pi}, \label{eq:Rdef}
\end{equation}
so that exactly $m$ charges, spaced by $a_0$, fit along the circumference. The particles can then be arranged in a perfect triangular lattice with $m$ equally spaced lattice planes occupying the region $x \in [-x_0, x_0]$, where
\begin{equation}
    x_0 = \frac{(m-1) d_0}{2}, \hspace{.2\linewidth}d_0 = \frac{\sqrt{3} a_0}{2}. \label{eq:x0lattice}
\end{equation}

Figure~\ref{fig:lattice} shows Monte Carlo samples at very low temperature, $\Gamma = 400$, where we observe the particles forming a triangular lattice. Due to the choice of the cylinder radius, we are able to adopt mixed boundary conditions: the positions of the particles in the lattice plane at $x = x_0$ are fixed, while those at the $x = -x_0$ boundary are free to fluctuate. This setup fixes the angular orientation of the cylinder (defining the position of the origin along the $y$-axis), enabling the separate analysis of fluctuations normal and parallel to the boundary. 

The right panel of Figure~\ref{fig:lattice} shows these fluctuations, where we observe that they increase logarithmically with distance from the boundary, as is characteristic of a two-dimensional crystal. At the left boundary, the fluctuations of particle positions normal to the boundary still vanish due to the strong influence of the external potential, $W = \pi x^2$.

\section{Monte Carlo simulations}
 \label{sec:MC}

We provide here a few practical details on the simulation method. The update scheme we use is similar to that in Ref.~\cite{cardoso2020boundary} for the disk geometry. To perform an update, we pick one of the $N$ particles uniformly at random and propose a move to a nearby location with a Gaussian probability distribution. The acceptance or rejection of the move is determined based on the Metropolis-Hastings scheme. The variance of the distribution is tuned to achieve a reasonable acceptance rate and ensure good thermalization properties.

Because of the form (\ref{eq:energy}) for the energy, each update can be performed in $O(N)$ time. However, compared to the disk geometry, the prefactor is significantly larger due to the need to evaluate the functions $\cos$ and $\cosh$ (see (\ref{intcyl})) instead of a second-degree polynomial. To mitigate this issue, we perform the updates in parallel, which results in a significant speedup for large particle numbers.

We verified thermalization by performing several independent runs and confirming that the results agree. We find that achieving thermalization is easier in the cylinder geometry than in the disk geometry. This is presumably because the energy is exactly minimized by a triangular lattice on the cylinder, whereas in the disk geometry, defects are unavoidable. In practice, we find that the improved thermalization offsets the slower update, justifying the choice of the cylinder geometry for this study.

In several cases, we also need to perform simulations where certain particles are fixed at specific positions. This is straightforward to achieve by initializing the selected particles at the desired locations and systematically rejecting updates involving those particles.

\section{Density profiles}
\label{sec:profiles}

\begin{figure}
\centering
\includegraphics[width=0.8\linewidth]{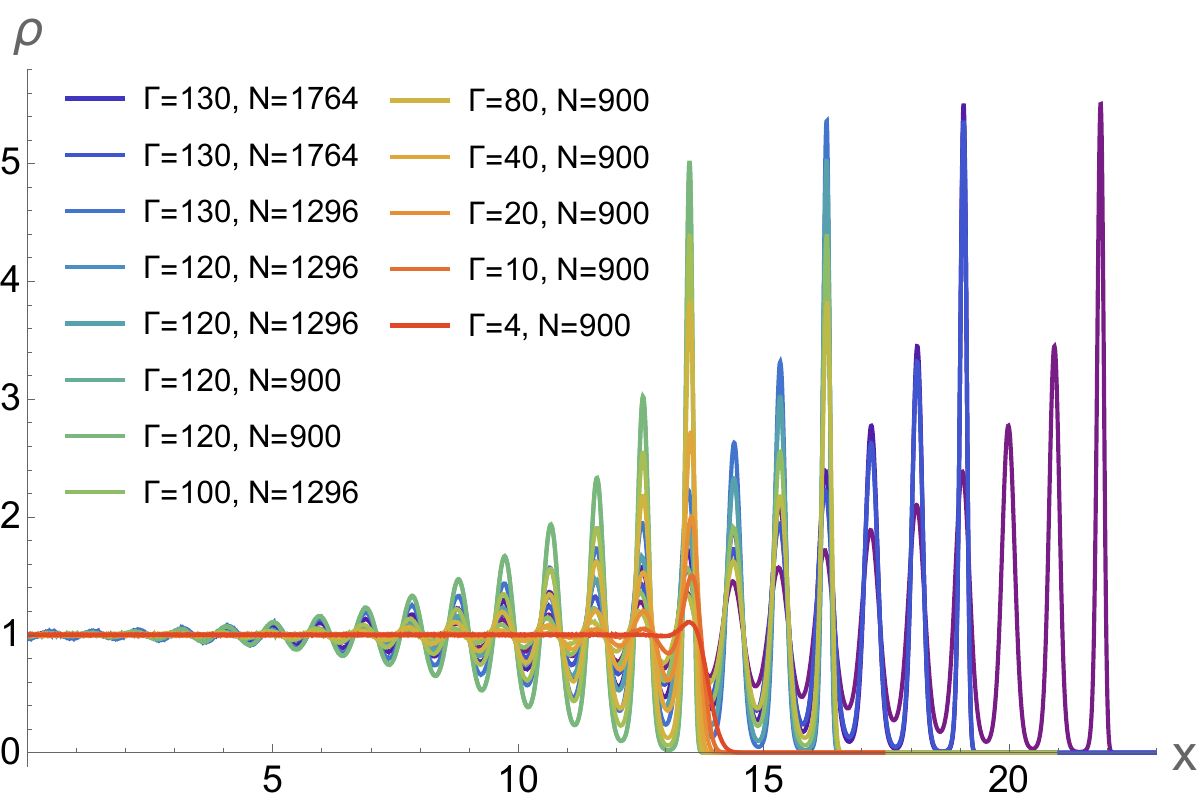}
\caption{Dependence of the density profile on temperature in the strong coupling fluid regime $2 < \Gamma < \Gamma_m$. In units where $\rho_0 = 1$, the wavelength of the edge oscillations is independent of $\Gamma$, while the damping length $\xi$ increases with $\Gamma$. As a result, the oscillations extend further into the bulk at lower temperatures.}
\label{fig:tdependence}
\end{figure}

\begin{figure}
\begin{subfigure}{.48\linewidth}
\centering
\includegraphics[width=\linewidth]{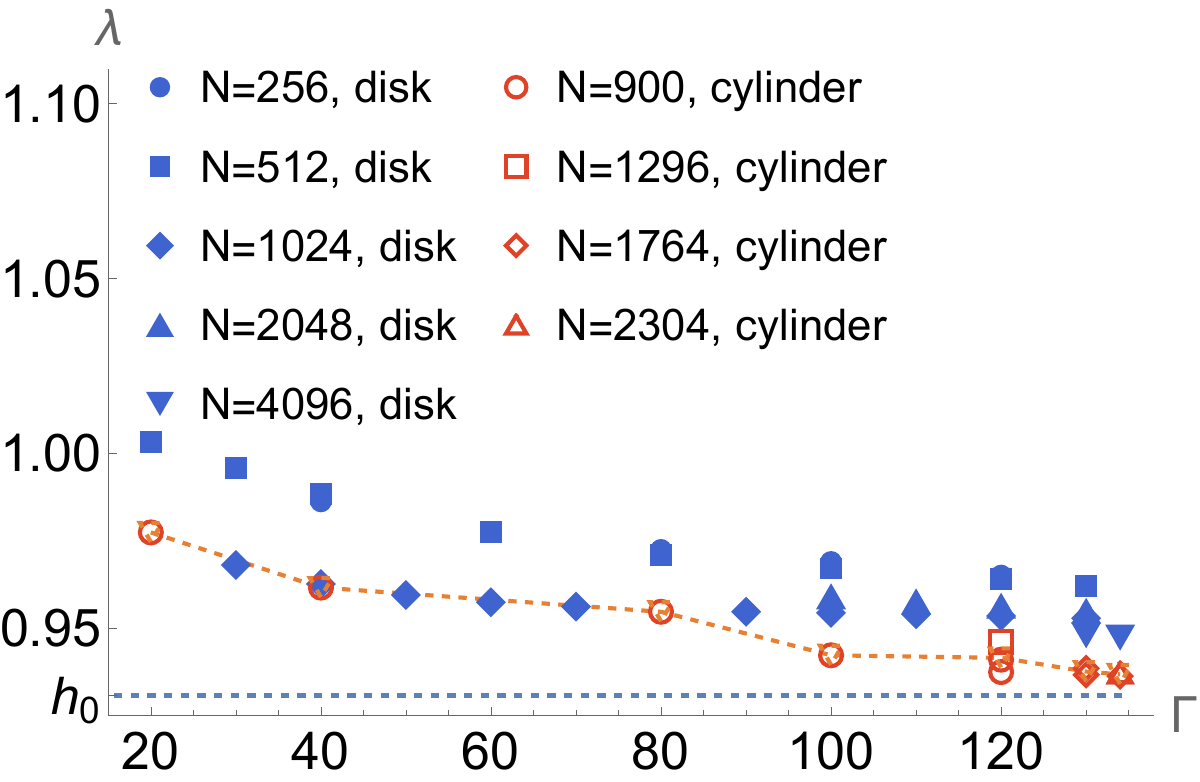}
\end{subfigure}
\hfill
\begin{subfigure}{.48\linewidth}
\centering
\includegraphics[width=\linewidth]{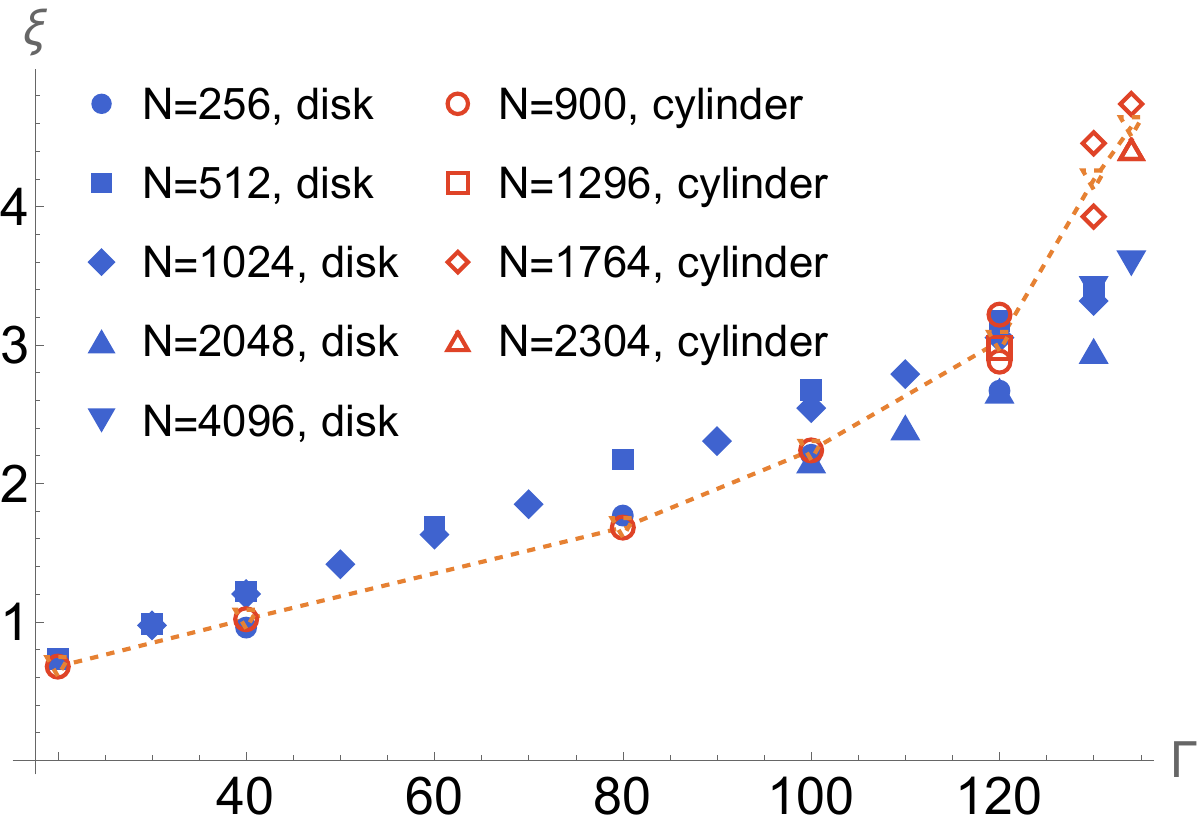}
\end{subfigure}
\caption{Temperature dependence of the wavelength $\lambda$ (left) and damping length $\xi$ (right) of boundary density oscillations. As $\Gamma$ approaches $\Gamma_m \approx 140$, the wavelength $\lambda$ converges to the triangular lattice value $d_0 \approx 0.93$, and the damping length $\xi$ increases. These features are more pronounced in the cylinder geometry compared to the disk data, although the two datasets show better agreement at large $N$.}
\label{fig:wavedamping}
\end{figure}

In the range $2 < \Gamma < \Gamma_m$, where the OCP is in the strong coupling fluid phase, the density profile exhibits damped oscillations near the boundaries, occurring at the scale of the inter-particle distance. These oscillations are shown in Fig.~\ref{fig:tdependence} for different values of $\Gamma$ and $N$. In our units, larger $N$ simulations correspond to larger droplet sizes; however, the wavelength and decay length of the oscillations remain unchanged. The oscillations become more pronounced as $\Gamma$ increases, corresponding to weaker damping or, equivalently, a larger decay length. We quantify this behavior by fitting the damped oscillations with the expression
\begin{equation}
    e^{-x/\xi} \sin(2\pi x/\lambda).\label{eq:fitoscillations}
\end{equation}
The extracted parameters are plotted in Fig.~\ref{fig:wavedamping} as a function of $\Gamma$.

These features cannot be explained by mean-field theory methods, which rely on the large $N$ expansion. In the limit of large $N$, with the droplet size $2x_0$ fixed, all oscillations are compressed into an overshoot singularity at the boundary \cite{zabrodin2006large}. In contrast, the damped oscillations are naturally explained in terms of the melting of a boundary crystalline layer \cite{cardoso2020boundary}. Specifically, the distance between peaks corresponds to the spacing between lattice planes in the triangular lattice, $d_0 \approx 0.93$, while the damping follows from the finite correlation length of lattice displacements in the melted phase.

This picture is even clearer in the cylindrical geometry, as shown in Fig.~\ref{fig:wavedamping}. As the temperature decreases toward the freezing transition, the wavelength rapidly converges to the triangular lattice value, and the correlation length exhibits a sharper increase, consistent with the expected divergence at the transition.

\subsection{Sensitivity to boundary conditions}
\label{sec:boundarycond}

\begin{figure}
\begin{subfigure}{.48\linewidth}
\centering
\includegraphics[width=\linewidth]{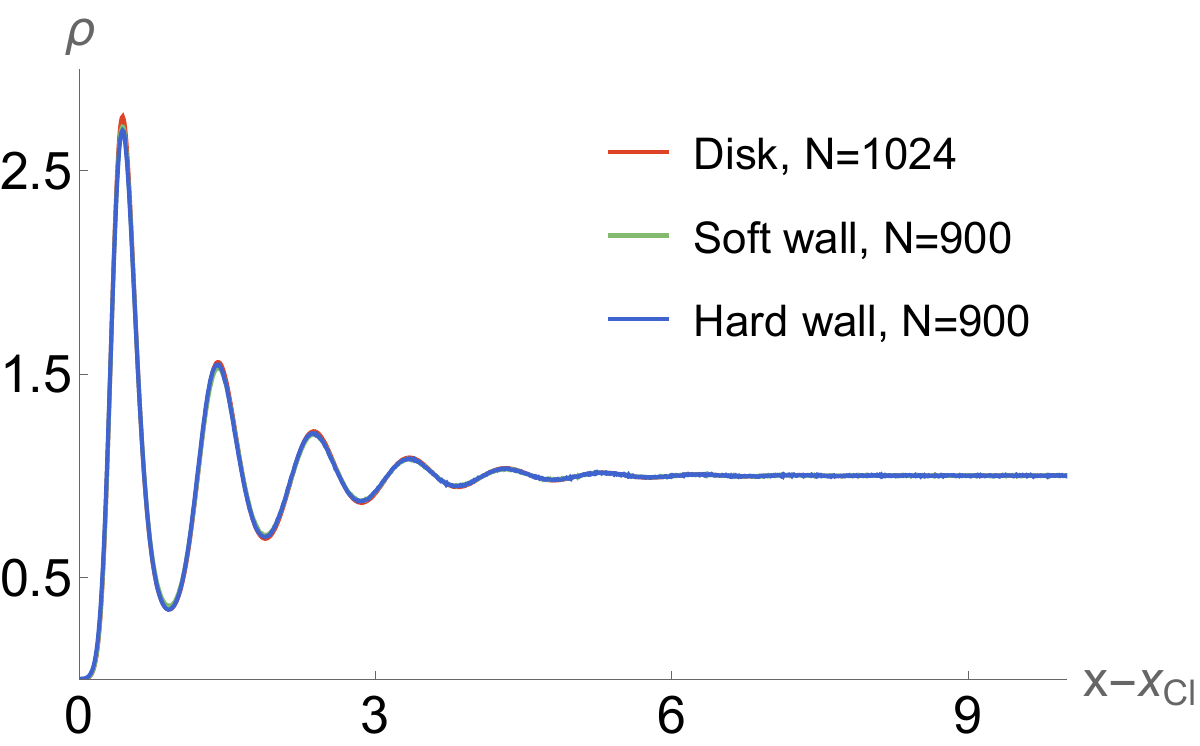}
\end{subfigure}
\hfill
\begin{subfigure}{.48\linewidth}
\centering
\includegraphics[width=\linewidth]{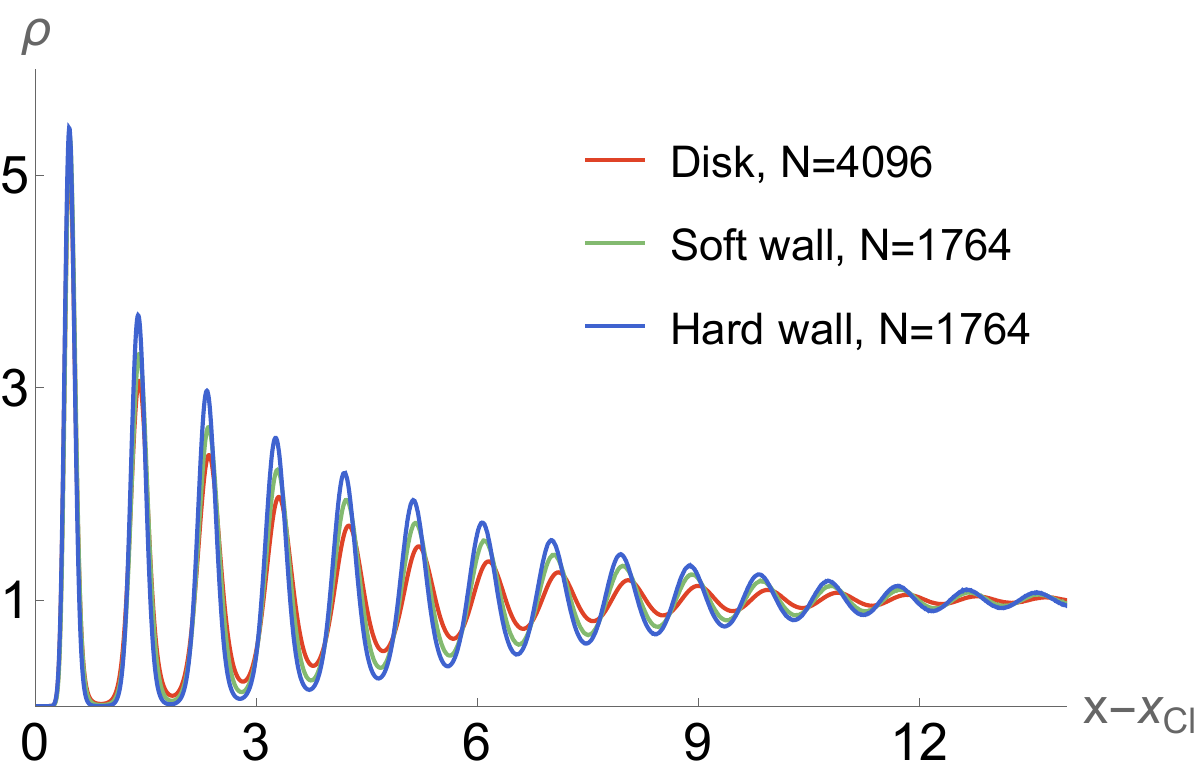}
\end{subfigure}
\caption{Comparison of density profiles with different boundary conditions at $\Gamma = 40$ (left) and $\Gamma = 130$ (right). The density is plotted as a function of the distance to the classical boundary $x_{\rm Cl}$, defined by the constant-density droplet. In the case of a hard wall, the origin is shifted by the distance $d_0$ from the fixed to the first fluctuating lattice plane. At small $\Gamma$, we see a perfect agreement between the different boundary conditions, reflecting the universal shape of the density profile due to screening in the bulk. Close to the transition, when many lattice planes are visible, we see that the oscillations in the disk become out of phase in the bulk due to the accumulated effects of geometric frustration. }
\label{fig:comparebc}
\end{figure}

In most of our simulations, we consider soft-wall boundary conditions: the positions of the charges are allowed to fluctuate across the entire plane, and the formation of the boundary arises solely from the minimization of energy for a finite number of particles in the presence of the external potential. Another choice, referred to as hard-wall boundary conditions, involves pinning a line of charges at the boundary $x = \pm x_0$. Mixed boundary conditions correspond to fixing the particles at one boundary only, as shown in Fig.~\ref{fig:lattice}. 

We find that the shape of the density profile is universal. For instance, in Fig.~\ref{fig:comparebc}, we compare the hard-wall, soft-wall, and disk (soft-wall) density profiles as a function of the distance to the boundary for two values of the inverse temperature $\Gamma$. The boundary position is defined as the classical boundary $x_{\rm Cl}$, determined by the constant-density droplet. For hard-wall boundary conditions, the origin is further shifted by the lattice plane spacing $d_0$, so that the density profiles are compared starting from the first fluctuating lattice plane. 

The surprising agreement between the density profiles demonstrates that the exact details of the boundary conditions are quickly screened by the plasma. This agreement can also be understood in terms of the equal values of $\sigma_x$, the amplitude of particle position fluctuations along the $x$ direction, as a function of the distance to the boundary. This behavior is shown in Fig.~\ref{fig:lattice} for the crystalline phase and in Fig.~\ref{fig:fluidboundary} for the fluid phase.

Comparing the soft- and hard-wall density profiles clarifies the phenomenon known as droplet squeezing for $\Gamma > 2$. It is known that the onset of boundary oscillations is accompanied by a shift in the boundary location in the soft-wall case. This squeezing effect is non-perturbative in the large $N$ limit and is related to the overshoot singularity at the boundary \cite{wiegmann2014anomalous}. The shift can be expressed as the mismatch between the boundaries of the constant-density fluid and the triangular lattice,
\begin{equation}
    \delta = \frac{1}{2}\left[\frac{m^2}{2\pi R} - (m-1)d_0\right] = \frac{d_0}{2},
\end{equation}
where we have used (\ref{eq:Rdef}) and (\ref{eq:x0lattice}). This provides a simple explanation for the value $\delta \approx 0.465$, which was argued in \cite{cardoso2020boundary} using screening arguments.

At large $\Gamma$ (Fig.~\ref{fig:comparebc}, right), where multiple lattice planes can be discerned, we observe that in the bulk, these planes become out of phase in the disk geometry due to geometric frustration effects. This results in lower accuracy when determining the wavelength and damping of oscillations, which serve as signatures of the thermodynamic properties of the plasma.

\subsection{Anisotropy and stripe order}
\label{sec:melting}

\begin{figure}
\begin{subfigure}{.48\linewidth}
\centering
\includegraphics[width=\linewidth]{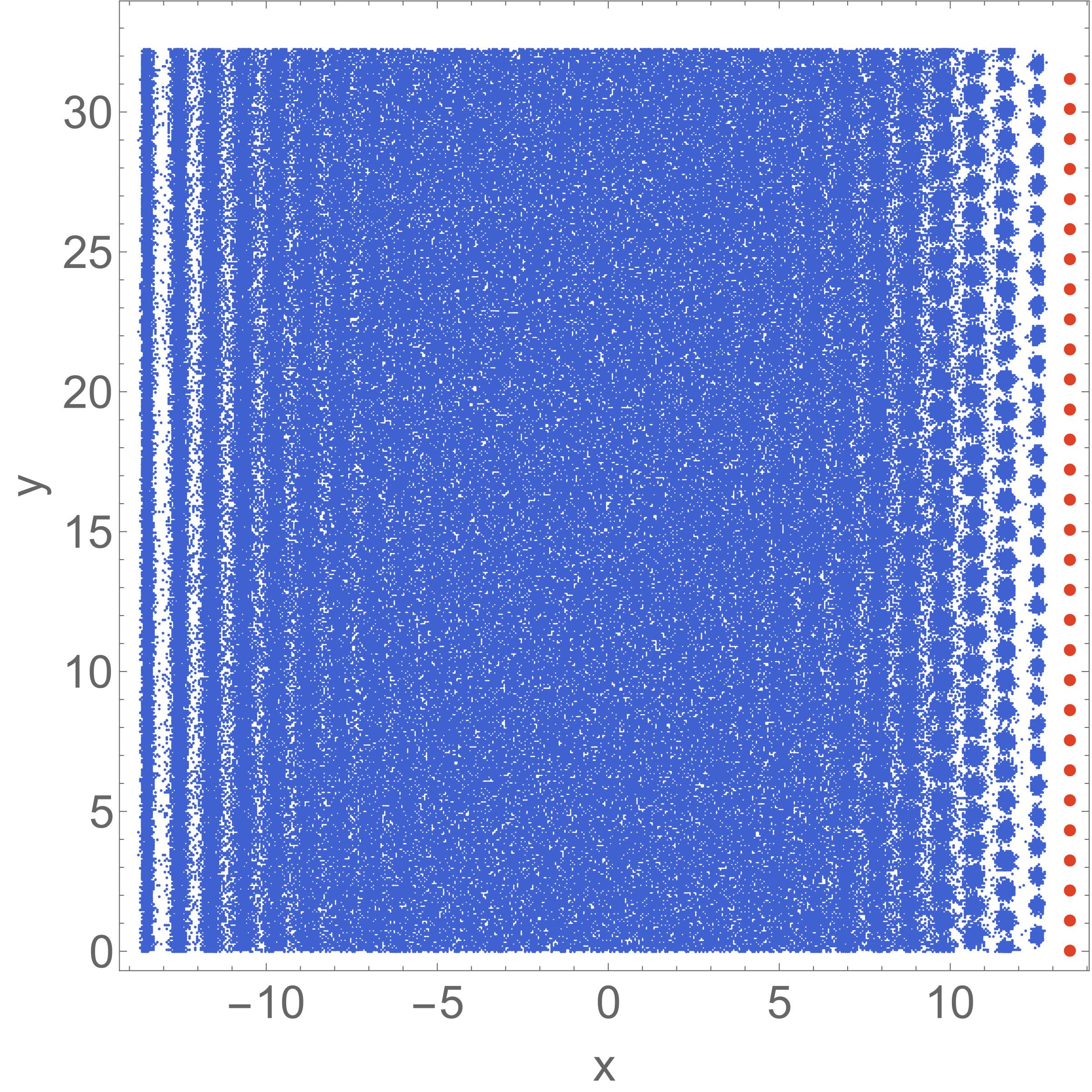}
\end{subfigure}
\hfill
\begin{subfigure}{.48\linewidth}
\centering
\includegraphics[width=\linewidth]{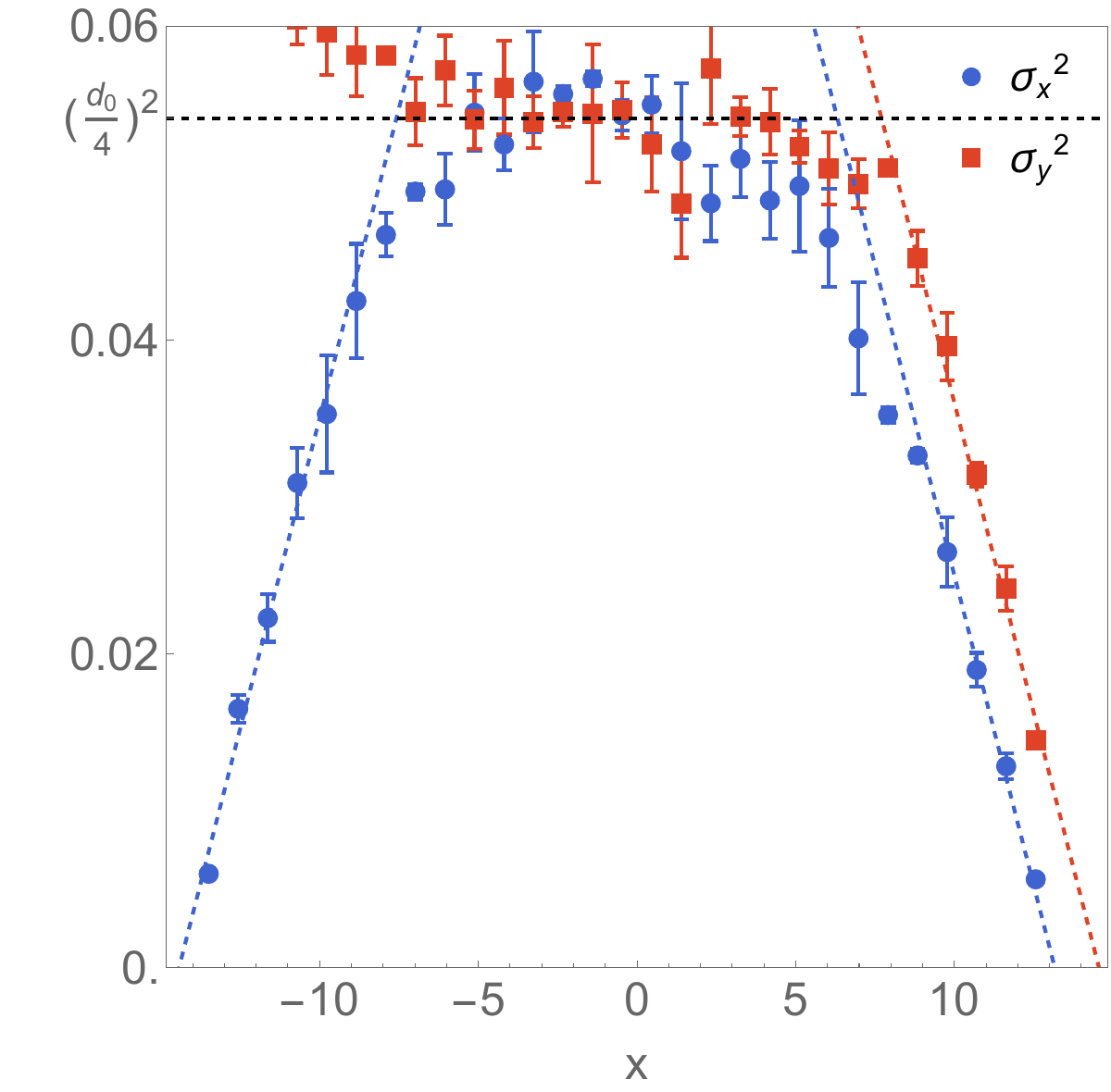}
\end{subfigure}
\caption{Two-dimensional density in the fluid phase. Left: A total of $512$ Monte Carlo samples for the plasma at $\Gamma = 120 < \Gamma_m$ with $N = 900$ particles and mixed boundary conditions. Right: Squared variance of particle positions along the $x$- and $y$-directions as a function of $x$. These variances vanish at the right boundary and increase linearly with distance (dashed lines). At the left boundary, only the fluctuations in the $x$-direction (normal to the boundary) vanish. When the variance reaches the order of $d_0/4$, the density becomes approximately homogeneous.}
\label{fig:fluidboundary}
\end{figure}

It is interesting to consider not only the density profiles but also the two-dimensional properties of the density. As shown in Fig.~\ref{fig:lattice}, the amplitude of particle position fluctuations is anisotropic in the crystalline phase. While hard-wall boundary conditions suppress fluctuations both normal and parallel to the boundary, soft-wall boundary conditions suppress only the normal fluctuations. In both cases, the amplitude of fluctuations increases logarithmically with the distance from the boundary, a characteristic feature of a two-dimensional crystal. Consequently, for a large enough sample, stripes will form near a soft-wall boundary.

This behavior is also observed in the fluid regime, as shown in Fig.~\ref{fig:fluidboundary}. Once again, hard-wall boundary conditions suppress fluctuations in both directions, while soft-wall boundary conditions suppress only the fluctuations normal to the boundary. Consequently, while the density profile remains independent of the choice of boundary conditions, the smearing of the density along the direction parallel to the boundary does depend on it. 

Finally, due to the melting of crystalline order, fluctuations grow faster in the fluid phase, with $\sigma^2$ increasing linearly from the boundary. The slope of this linear increase is related to the damping length of the density oscillations.

\section{Topological defects}
\label{sec:topdef}

In addition to fluctuations around the triangular lattice sites, it is insightful to quantify the density of topological defects at low temperatures. These defects can be identified through the following procedure. Given a snapshot of the particle configurations, we construct the Delaunay triangulation to determine the ideal triangular lattice formed by the particles. Each particle is then assigned a coordination number $CN$, representing the number of particles connected to it by an edge. In a perfect triangular lattice, all points have $CN = 6$, while deviations from this are known as disclinations. 

\begin{figure}
\centering
\includegraphics[width=\linewidth]{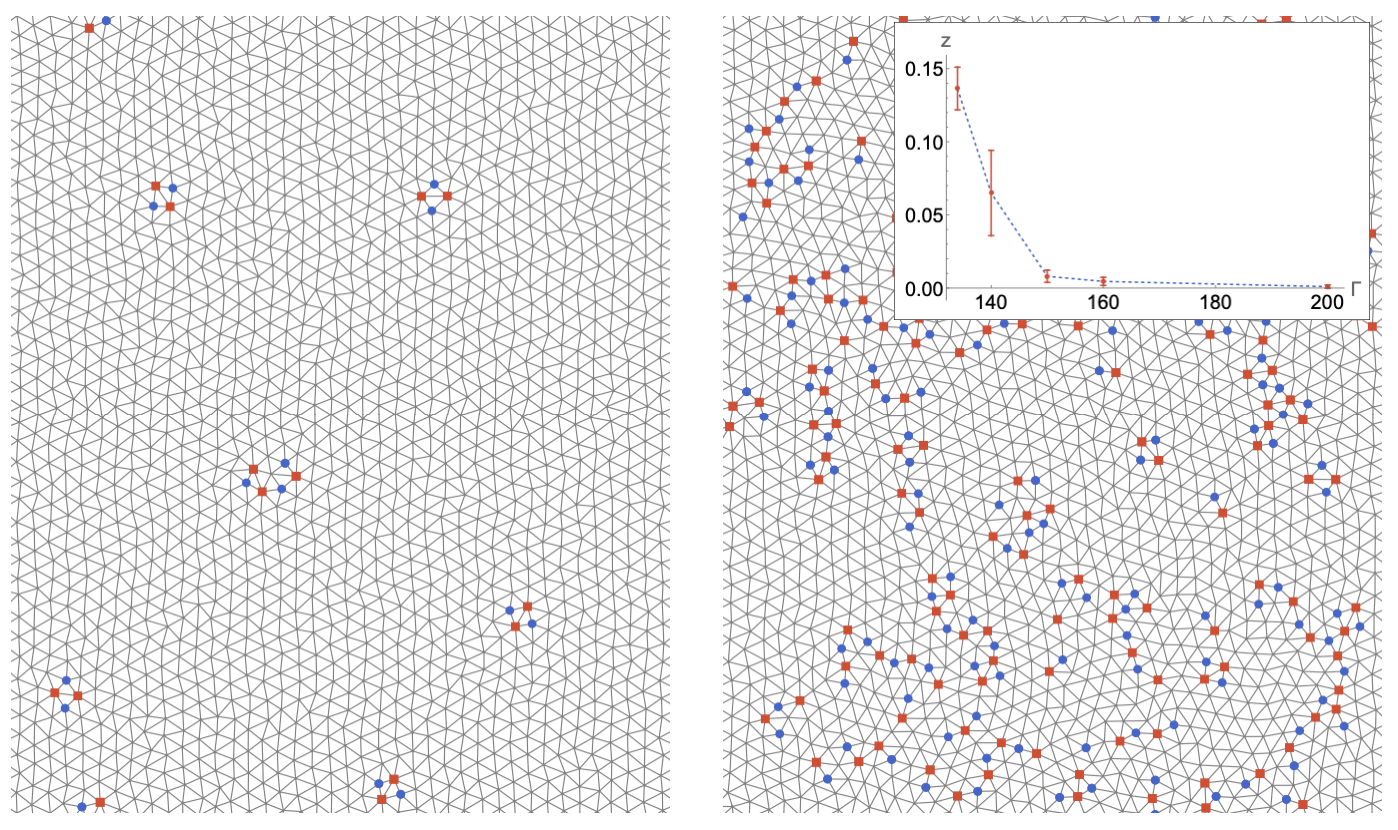}
\caption{At high $\Gamma$, the particles form a triangular lattice with a few disclination defects (example shown on the left for $\Gamma = 150, N=1764$), while at lower $\Gamma$, the density $z$ of defects increases rapidly (example shown on the right for $\Gamma = 130, N=1764$). The most common defects are the points of coordination five and seven, which are shown in red and blue, respectively. The inset shows the dependence of $z$ on $\Gamma$.}
\label{fig:topdefects}
\end{figure}

Above the transition ($\Gamma > \Gamma_m$), we find few, localized disclinations that form tightly bound dipoles of $CN = 5$ and $CN = 7$ disclinations, as shown in Fig.~\ref{fig:topdefects}. At higher temperatures, disclinations become more common. Below the transition, however, disclinations are meaningful only for individual snapshots, as they are strongly averaged out by lattice fluctuations. We extract the average density $z$ of dislocations, which is plotted as a function of $\Gamma$ in Fig.~\ref{fig:topdefects}. 

The density of dislocations is exponentially suppressed above the transition ($\Gamma_m \approx 140$), consistent with the high energy cost of creating topological defects in the crystal phase. We do not observe a strong divergence in the density of defects near the transition, which would favor the weakly first-order melting scenario. However, as is well-known, the divergence in defect concentration characteristic of the BKTHNY transition can be elusive in numerical simulations \cite{kosterlitz2016kosterlitz}.

\section{Density correlation functions}
\label{sec:correlations}

\begin{figure}
  \centering
    \includegraphics[width=0.7\textwidth]{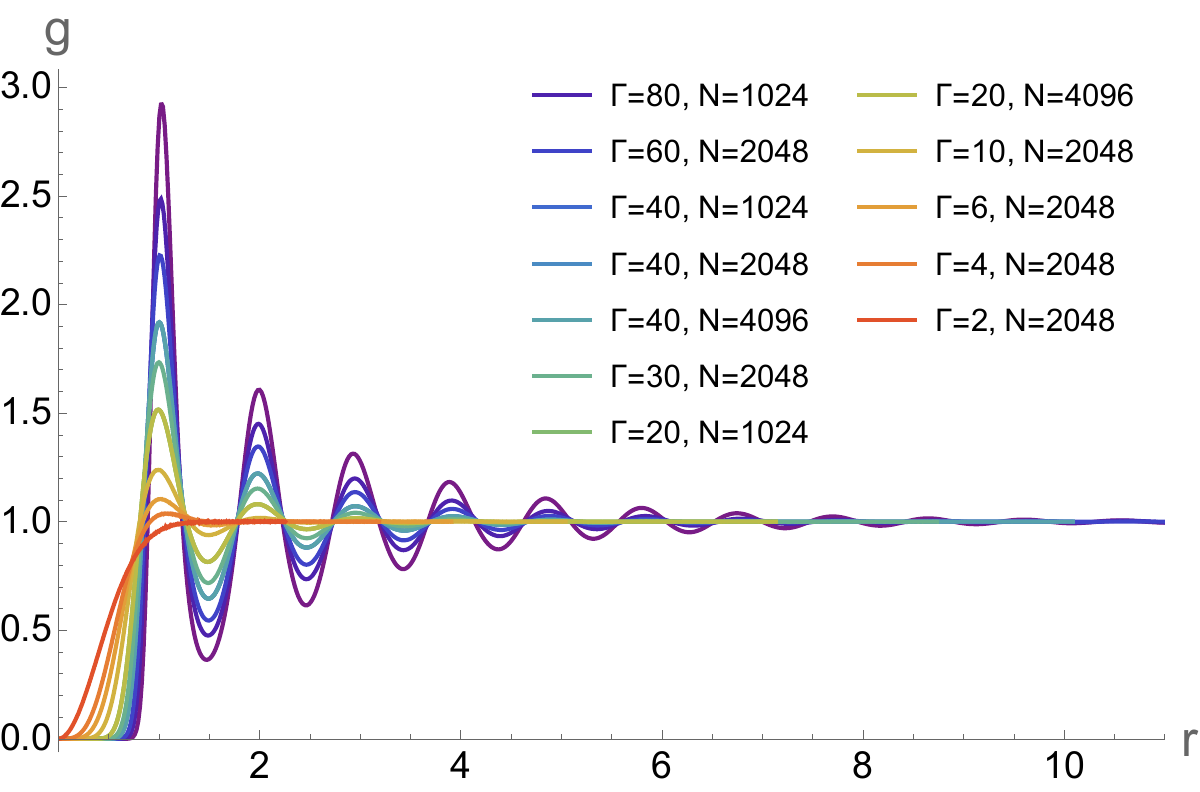}
  \caption{Radial distribution function $g(r)$ for different values of $\Gamma$ and $N$. The peaks of $g(r)$ indicate short-range ordering in the system, becoming more pronounced as $\Gamma$ increases, reflecting stronger correlations at lower temperatures.}
  \label{fig:gr}
\end{figure}

The damped oscillations we discussed also appear in the bulk density correlation functions \cite{fertig1987hypernetted}. In particular, we study the radial distribution function,
\begin{equation}
    g(|\mathbf{x} - \mathbf{x'}|) = \frac{\langle\hat{\rho}(\mathbf{x})\hat{\rho}(\mathbf{x'})\rangle}{\rho(\mathbf{x})\rho(\mathbf{x'})}, \label{eq:hrhocorrelation}
\end{equation}
where $g$ is translationally invariant in the bulk. We evaluate $g(r)$ from the simulations by considering correlations between points deep in the bulk. The results are shown in Fig.~\ref{fig:gr}. 

The connection between the boundary density profile and the correlation functions can be understood in terms of the phase-field crystal (PFC) free energy,
\begin{equation}
    {\cal F}[\psi] = \int d^2x \left\{\frac{a}{2}\psi^2 + \frac{b}{2}\psi(\Delta + q_0^2)^2\psi + u\frac{\psi^4}{4} + V\psi\right\}. \label{eq:FPFC}
\end{equation}
Here, the field $\psi = \frac{\rho - \rho_0}{\rho_0}$ corresponds to the normalized density fluctuations away from the homogeneous fluid value $\rho_0 = 1$. The phenomenological parameters $a$, $b$, $u$, and $q_0$ are either fitted or computed from a microscopic formulation of the liquid \cite{elder2004}. For a given boundary external potential $V$, the density profile $1 + \psi$ is determined by the saddle-point equation,
\begin{equation}
    \left[a + b(\Delta + q_0^2)^2\right]\psi + u\psi^3 + V = 0.
\end{equation}
Deep in the bulk, the non-linear term and the potential term can be neglected, leading to density oscillations with wavevector
\begin{equation}
    k^2 = q_0^2 \pm \sqrt{-\frac{a}{b}}.
\end{equation}
These oscillations may be damped or undamped depending on the relative sign of $a$ and $b$.

The PFC parameters are related to the density correlation functions of the bulk fluid. Expanding the free energy around the homogeneous bulk density $\rho_0$ gives
\begin{equation}
    {\cal F}[\rho_0 + \delta\rho] = \frac{1}{2} \int d^2x\, d^2x'\, \delta\rho(\mathbf{x}) \left[1 - c_0(|\mathbf{x} - \mathbf{x'}|)\right] \delta\rho(\mathbf{x'}), \label{eq:Fdeltarho}
\end{equation}
where $c_0(r)$ is the direct correlation function, defined as the second derivative of the excess free energy \cite{hansen2013theory}. The Ornstein-Zernike relation,
\begin{equation}
    h_0(k) = \frac{c_0(k)}{1 - c_0(k)},
\end{equation}
relates it to the total correlation function $h_0(r) = g(r) - 1$. In the PFC approximation,
\begin{equation}
    h_0(k) = -1 + \frac{1}{a + b(q_0^2 - k^2)^2},
\end{equation}
the quadratic part of the free energy (\ref{eq:Fdeltarho}) reduces to the quadratic part of PFC free energy (\ref{eq:FPFC}). The PFC model approximates the local structure of particle distributions in the fluid by focusing on a single peak at $k=q_0$ of the total correlation function $h_0(k)$. Equivalently, this corresponds to a peak of the static structure factor,
\begin{equation}
    S(k) = 1 + \int d^2r\, e^{i\mathbf{k} \cdot \mathbf{r}} g(r).
\end{equation}

\begin{figure}
\begin{subfigure}{.48\linewidth}
\centering
\includegraphics[width=\linewidth]{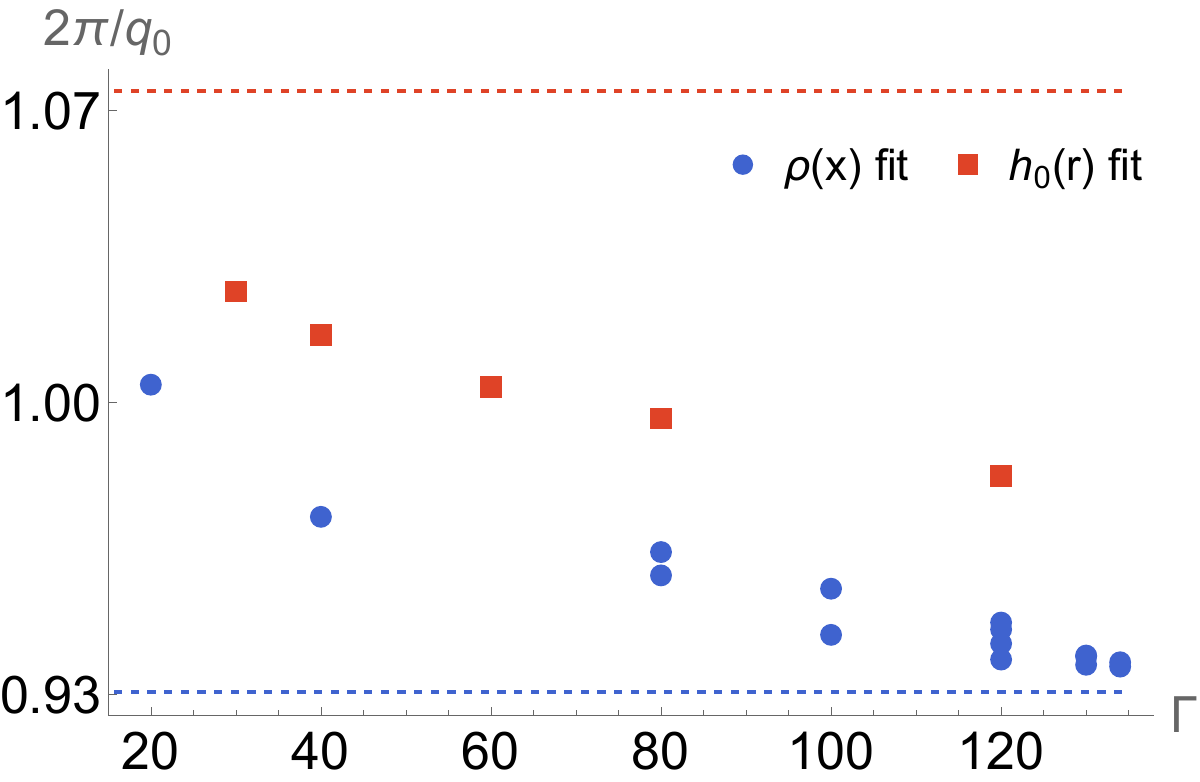}
\end{subfigure}
\hfill
\begin{subfigure}{.48\linewidth}
\centering
\includegraphics[width=\linewidth]{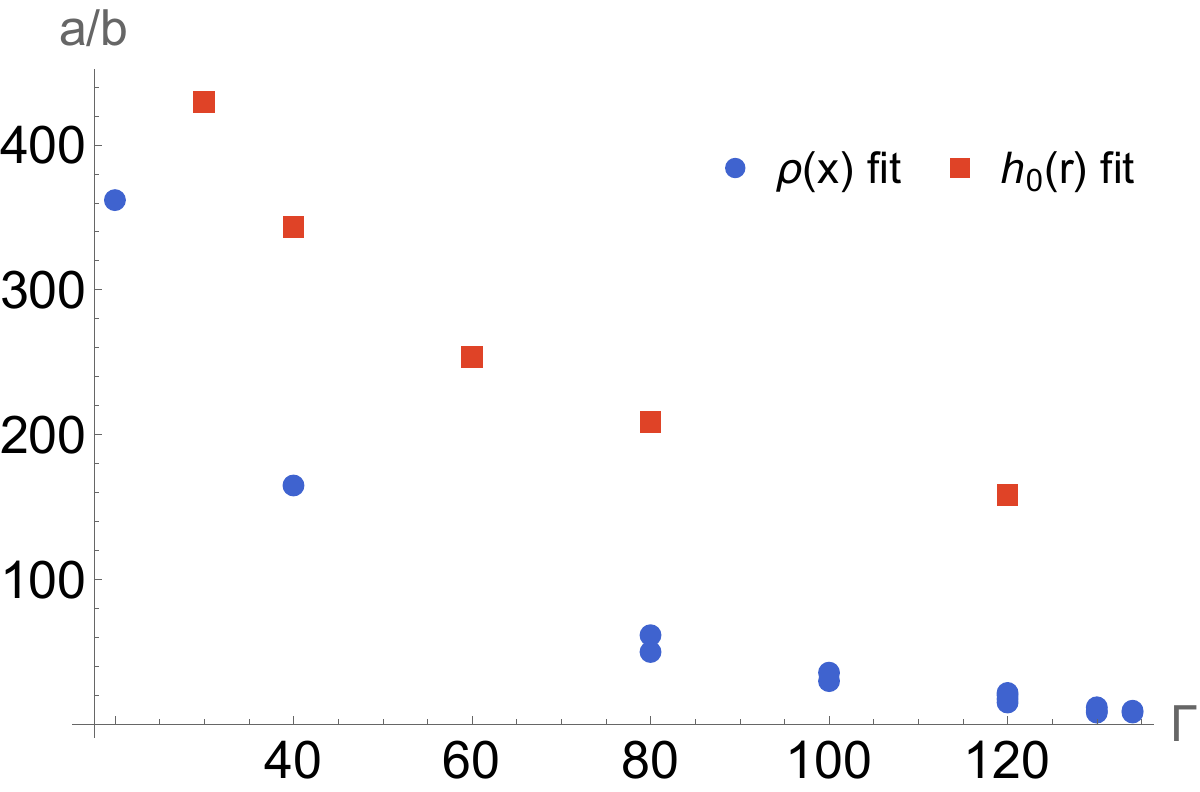}
\end{subfigure}
\caption{Fit of PFC parameters determined from the total correlation function $h_0(r)$ and the boundary density profiles $\rho(x)$. The reference values $a_0 \approx 1.07$ and $d_0 \approx 0.93$ are shown for comparison in the $q_0$ fit.}
\label{fig:q0ab}
\end{figure}

This formulation also relates the damped oscillations to the crystallization transition. The peak of the static structure factor at $q_0$ corresponds to the leading reciprocal lattice vector of the crystalline phase, while $1/a$ determines the peak height and $b$ specifies the curvature of $S(k)$ at the peak. The ratio $a/b$ is positive in the fluid phase, leading to damped oscillations in the density profile, and vanishes at the transition $\Gamma \to \Gamma_m$, corresponding to an infinite damping length. 

We verify this correspondence by fitting the parameters $q_0$ and $a/b$ from the density profile and the radial correlation function, as shown in Fig.~\ref{fig:q0ab}. The results demonstrate general agreement, though the actual values of $q_0$ and $a/b$ differ by approximately five and twenty percent, respectively.

\begin{figure}
  \centering
    \includegraphics[width=.9\textwidth]{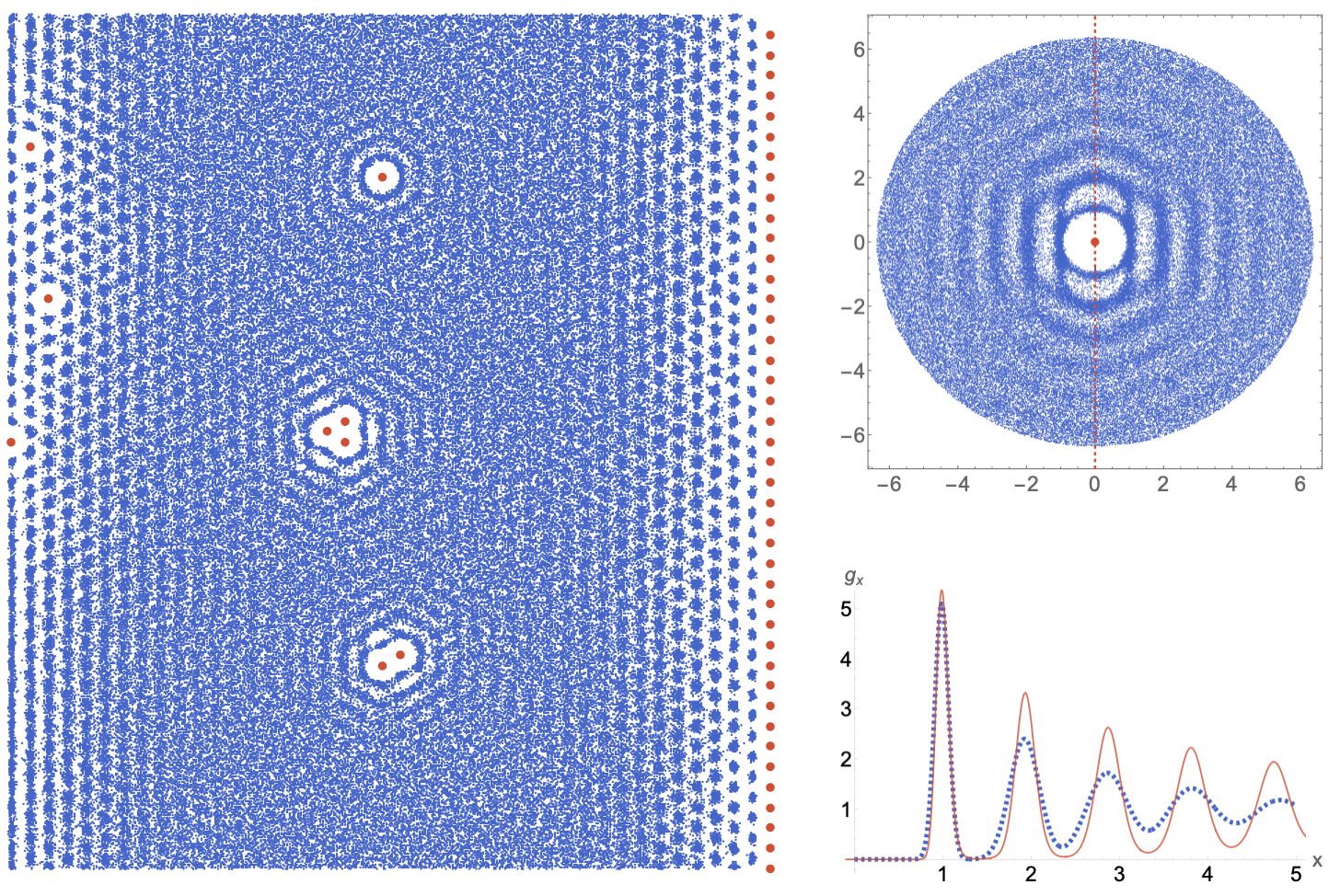}
    \caption{Left: Snapshots with fixed particles (red). The density of particles around a single fixed particle defines the correlation function $g(r)$. Similarly, higher-point correlation functions are related to the density around multiple fixed particles, which can be positioned at the sites of a triangular lattice.  Right, top: The definition of $g(r)$ can be modified by fixing the orientation of the snapshots, aligning the particle closest to the test particle along the $y$-axis. Right, bottom: The resulting density along the $x$-axis defines the oriented correlation function $g_x(x)$ at $\Gamma = 130$ (blue). The wavelength of oscillations shows improved agreement with the edge density profile at the same temperature (red), which suggests that an anisotropic generalization of the PFC action is required.}
\label{fig:fixedrhog}
\end{figure}

A somewhat better agreement can be achieved by defining an oriented generalization of the correlation function, illustrated in Fig.~\ref{fig:fixedrhog}. The key idea is that, while $g(r)$ represents the density around a fixed test particle, the boundary density profile corresponds to the density near a fixed line of test particles. In simpler terms, the boundary perturbation introduces an orientation. 

This distinction can be formalized by defining an oriented correlation function. Specifically, the orientation of each snapshot is fixed by aligning the vector pointing to the closest particle to the origin. Averaging over these oriented snapshots defines the oriented correlation function $g_x(x)$. As shown in Fig.~\ref{fig:fixedrhog}, we observed that the wavelength of the oscillations in this correlation function agrees much more closely with the wavelength extracted from the density profile. This suggests that an anisotropic generalization of the PFC action is required to more accurately match the properties of the density profile with the correlation functions \cite{elder2004}.

\section{Discussion}
\label{sec:discussion}

We presented new Monte Carlo data for the two-dimensional one-component plasma (2D OCP) confined to a cylindrical geometry, focusing on the density profile, fluctuations, and their connection to bulk correlation functions and the phase diagram. Using Monte Carlo simulations, we examined the behavior of the system across a range of temperatures, with particular emphasis on the strongly coupled fluid phase near the Wigner crystallization transition. The cylindrical geometry allowed us to eliminate geometric frustration and the accompanying disclinations commonly observed in planar disk geometries at finite particle numbers.

At low temperatures, the local structure of the 2D OCP is that of a triangular lattice. Using hard-wall boundary conditions to fix the orientation around the cylinder, we found that particle position fluctuations increase logarithmically with the distance from the boundary in the crystalline phase. By triangulating the Monte Carlo samples, we captured the exponential suppression of topological defects for $\Gamma > 140$. However, within the limitations of our numerical approach, we did not observe the divergence in defect density characteristic of the BKTHNY transition, supporting the weakly first-order melting paradigm.

In the fluid phase, fluctuations of particle positions along the direction parallel to the boundary tend to smear the density, resulting in a striped particle density. This is reflected in the density profile oscillations, which have a wavelength determined by the triangular lattice spacing. The effective squeezing of the OCP droplet can be explained in a similar way. These oscillations are damped, with the damping length increasing sharply near the freezing transition. We found that these features are more pronounced in the cylinder geometry, due to the absence of geometric frustration and the clear separation of directions parallel and normal to the boundary.

We verified that these conclusions are generic by comparing different boundary conditions and geometries. Additionally, we introduced a phenomenological phase-field crystal (PFC) model to summarize our findings. This model captures the relationship between the melting transition, bulk correlation functions, and boundary density oscillations. It also provides a framework for connecting density profiles to the phase diagram through the leading peak of the static structure factor. 

Finally, we highlighted that, while the radial total correlation function is closely related to the fluid density around a test particle, the density profile at a planar wall is more closely associated with oriented correlation functions. We thus expect that an extension of the PFC model incorporating orientational correlations will enable a more precise calculation of the density profile.

\section*{Acknowledgments}

GC acknowledges support from a TDLI postdoctoral fellowship. AGA's work was supported by the National Science Foundation under Grant NSF DMR--2116767.

\clearpage

\appendix

\section*{References}

\bibliography{coulomb-droplet}

\bibliographystyle{unsrt}

\end{document}